\documentclass[10pt,reprint,amsmath,amssymb,aps,prb,twocolumn]{revtex4-2}
\usepackage{preamble}

\usepackage{notes2bib}

\begin{document}

\title{Trion liquid and its photoemission signatures}

\author{Noam Ophir}
\affiliation{Department of Physics, Technion, Haifa, 3200003, Israel} 

\author{Anna Keselman}
\affiliation{Department of Physics, Technion, Haifa, 3200003, Israel}
\affiliation{The Helen Diller Quantum Center, Technion, Haifa, 3200003, Israel}


\begin{abstract}
We study the formation of a trion liquid in doped low-dimensional semiconductors with strong electron–hole interactions and analyze its signatures in angle-resolved photoemission spectroscopy (ARPES). We show that this strongly correlated state of matter forms naturally in the vicinity of the phase boundary between a normal band insulator and an excitonic insulator upon doping. By studying the photoemission spectrum, we show that a partially occupied trion band gives rise to in-gap features in the ARPES spectrum with vanishing spectral weight at the Fermi energy.
We demonstrate our findings using a 1D microscopic model employing exact, unbiased, matrix product state (MPS)-based calculations.
\end{abstract}

\maketitle

\section{Introduction}
In low-dimensional semiconductors, strong Coulomb interactions can drive the formation of multi-particle bound states that dramatically alter the behavior of the system. A prominent example is the formation of excitons -- bound electron-hole pairs. When the exciton binding energy becomes comparable to the band gap, the system may undergo a transition into an excitonic insulator (EI), a unique phase of matter in which excitons, which are bosonic quasiparticles, condense~\cite{JeromeKohn1967}. Several candidate EI materials have been proposed in recent years, both in quasi-one-dimensional systems~\cite{WakisakiTakagi2009,LuTakagi2017,TaPdTe_Huang2024,Hossain2025TopEI} and in two-dimensional transition-metal-dichalcogenide (TMD) double layers~\cite{Wang2019EI,Ma2021EI,Chen2022EI,Gu2022EI,qi2025competition}.

Another intriguing scenario is the formation of larger composite objects, such as trions -- bound states of an exciton with an additional electron or hole. 
These charge-imbalanced fermionic quasiparticles have been extensively studied in optically pumped systems in the past~\cite{Kheng1993,FinkelsteinBarJosesph1995,Mak2013Trions,Ross2013Trions,wang2018colloquium}, while recent experimental studies have demonstrated that they can also form in equilibrium. Specifically, it was demonstrated that TMD bilayers in which the electron-to-hole density ratio is tuned to 1:2 (or 2:1) can realize trion liquids -- Fermi liquids of trions, motivating theoretical proposals for realizing novel correlated phases of matter in such systems~\cite{ZerbaKnap2024,DaiFu2024}.

In this work, we consider an alternative route to the formation of trions in equilibrium, which does not rely on fine-tuning the electron-to-hole density ratio. We show that doping a semiconductor with strong exciton binding, proximate to but outside the EI regime, can drive spontaneous trion formation. This scenario relies on a sizable binding energy between doped charge carriers and excitons: a doped carrier can induce the spontaneous formation of an exciton and subsequently bind to it, thereby lowering the total energy of the system.

Our work is motivated predominantly by the quasi-one-dimensional family of materials \TNSSE, in which substitution of selenium by sulfur increases the band gap, driving the system from the EI regime toward a conventional semiconductor~\cite{lu_zero-gap_2017,Volkov2021,chen_anomalous_2023}.
In this setting, strong electron-hole attraction is expected to also persist on the semiconducting side of the phase diagram, providing favorable conditions for equilibrium trions near the critical S/Se ratio. Furthermore, recent ARPES studies of \TNS\ find indications of trion formation~\cite{nitzav2025trion},
motivating a broader theoretical framework for equilibrium trion liquids and their photoemission signatures.

Here, we begin by establishing the necessary ingredients for the formation of an equilibrium trion liquid upon doping a narrow-gap semiconductor with strong electron-hole attraction. We then explicitly demonstrate this mechanism in a strongly interacting microscopic one-dimensional model using unbiased matrix-product-state (MPS)-based simulations, and show that a partially occupied negatively charged trion band produces characteristic in-gap spectral features in ARPES. While the model is motivated by quasi-one-dimensional materials, the mechanism identified here is not tied to a specific compound and applies more broadly.

\begin{figure}[t]
    \centering
    \includegraphics[width=\linewidth]{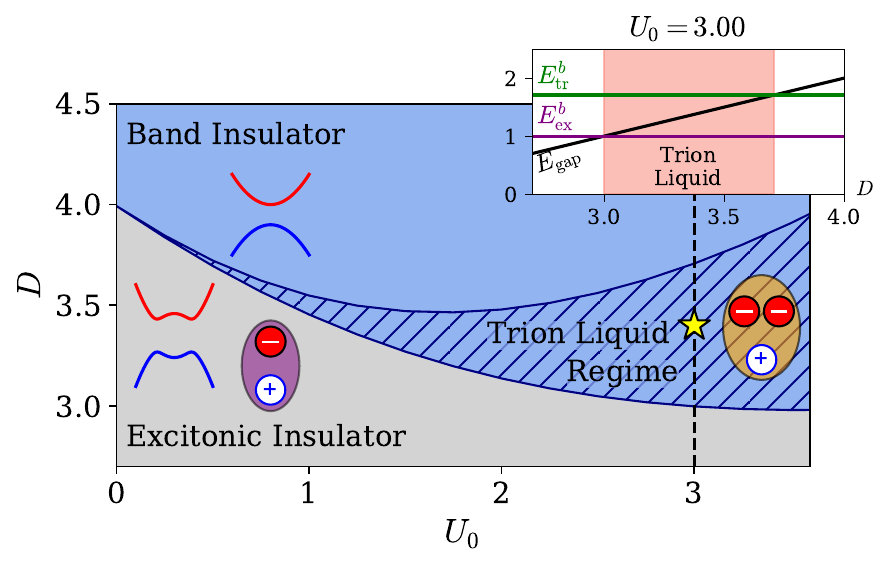}
    \caption{Phase diagram obtained for the model in Eq.~\eqref{eq:H2c1f}. The light blue and grey regions correspond to the band insulator (BI) and excitonic insulator (EI) phases obtained at charge neutrality, respectively. The dashed region, 
    indicates the parameter regime in which a trion liquid forms upon doping. 
    Model parameters used are $t_f/t_c=1,U_1/U_0=1/3$. The inset shows a cut across the phase diagram at fixed $U_0$ varying $D$, along the dashed line. The trion liquid phase is obtained in the regime $\Eex<\Eg<\Etr$ upon doping. The yellow star marks the point $D=3.4$, $U=3.0$ at which Figs. \ref{fig:energetics}(b), \ref{fig:model}(b), and \ref{fig:ARPES} are obtained. At this point, the binding energies are equal to $\Eex=1.00$, $\Eexe=0.71$ and the gap is $\Eg=1.40$. All energies are given in units of $t_c$.}
    \label{fig:PD}
\end{figure}

\section{Trion liquid phase}
We consider a narrow-gap semiconductor, with a single-particle gap $\Eg$, and exciton binding energy $\Eex$.
At charge neutrality, and in the regime $\Eex<\Eg$, the system is a simple band insulator. 
Upon doping, by either electrons or holes, formation of trions may be possible, due to the extra binding energy of the charge carrier to the exciton. Below we focus on electron doping, and denote the binding energy between an exciton and an electron by $\Eexe$. (The case of hole doping is analogous and we comment on it below.)
If the total trion binding energy, $\Etr
= \Eex+\Eexe$, is larger than the gap, $\Etr>\Eg$, the presence of the additional electron carrier can lead to a spontaneous generation of an exciton that binds to it, yielding a stable trion.
The regime $\Eex<\Eg<\Etr$, namely in proximity to, but outside, the EI phase (see Fig.~\ref{fig:PD}), is in the focus of this study.

\begin{figure}[t]
    \centering
    \llap{
    \parbox{0pt}{\vspace{0pt}\hspace{0pt}\footnotesize{(a)}}
    }
    \vspace{-10pt}
    \begin{minipage}{0.3\linewidth}
        \centering
        \includegraphics[width=\textwidth]{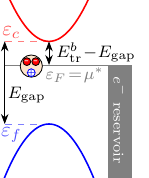}
    \end{minipage}
    \llap{
    \parbox{0pt}{\vspace{-106pt}\hspace{0pt}\footnotesize{(b)}}
    }
    \begin{minipage}{0.6\linewidth}
    \vspace{0.5cm}
        \centering
        \includegraphics[width=\textwidth]{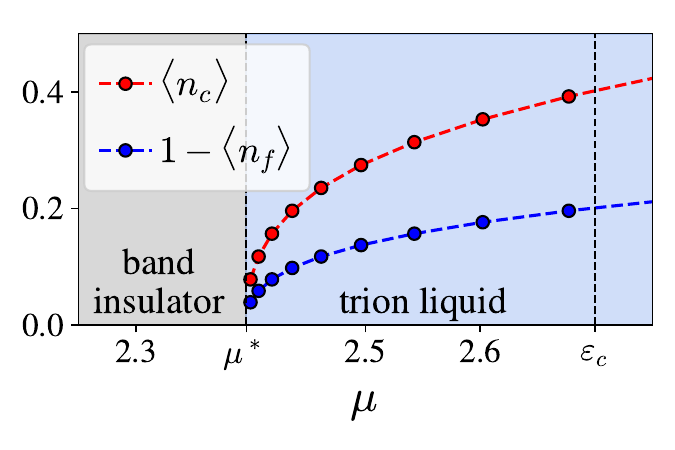}
    \end{minipage}
   
    \caption{(a) Schematic representation of the semiconducting system coupled to a charge reservoir, which fixes the Fermi energy, $\Ef$. 
    In the trion liquid regime, due to the energy gain upon trion formation, tunneling of electrons into the system is possible for $\mu>\mu^*=\Ec-(\Etr-\Eg)$ (see text).
    (b) Density of electrons and holes as a function of the chemical potential obtained numerically for the model in Eq.~\eqref{eq:H2c1f} with parameters corresponding to the point marked by a star in Fig.~\ref{fig:PD} and $V/U_0=1/6$ for an $N=51$-long chain. When the chemical potential crosses $\mu^*$, the density of electrons and holes starts increasing in a ratio of 2:1 as expected for the formation of trions. 
    }
    \label{fig:energetics}
\end{figure}

We consider an experimental setup in which the system is coupled to a reservoir that sets the Fermi energy and thereby determines the electronic density~\cite{nitzav2025trion}. 
In a non-interacting case, an electron can tunnel into the system, driving it away from charge neutrality, when the Fermi energy is above the bottom of the conduction band (which we denote by $\Ec$). However, when there is non-zero binding energy between the electron and an exciton, $\Eexe>0$, this condition is modified as illustrated in Fig.~\ref{fig:energetics}(a).
For the tunneling process of an electron to be energetically favorable, the single-particle cost $\Ec-\mu$ has to be compensated by the energy gain due to the formation of the trion. The latter is given by the energy gained upon exciton formation, $\Eex-\Eg$, plus the binding energy of an exciton to the electron, which sums up to $\Etr-\Eg$. We denote the chemical potential at this point by 
\begin{equation}\label{eq:mu_c}
\mu^*=\Ec-(\Etr-\Eg).    
\end{equation}
In Fig.~\ref{fig:energetics}(b) we plot the density of electrons and holes as function of the chemical potential for the 1D model, Eq.~\eqref{eq:H2c1f}, obtained numerically. As argued above, the density of conduction electrons becomes non-zero at $\mu=\mu^*<\Ec$. Moreover, the increase in conduction electron density occurs simultaneously with a rise in valence band hole density, maintaining a 2:1 electron-to-hole ratio as expected for trions.

The considerations and analysis above are analogous for the case of positively-charged trions that can form upon hole doping. In this case, the hole tunnels into the valence band when $\mu-\Ev<\Etr-\Eg$, where $\Etr$ now is given by the sum of the exciton binding, $\Eex$, and the binding of the exciton to the hole, $\Eexh$.

A distinguishing property of the trion liquid phase is that while it is a gapless metallic state, single particles are gapped. This is demonstrated explicitly in the static correlations plotted in Fig.~\ref{fig:model}(b). At sufficiently high charge-carrier density, the system is likely to become a two-component Fermi liquid in which trions and single electrons (holes) coexist. Here, we stay in the regime of low doping, in which all available carriers form bound trions.

\section{Microscopic model}
\begin{figure}
    \centering
    \llap{
        \parbox{0pt}{\vspace{0pt}\hspace{0pt}\footnotesize{(a)}}
        }
    \begin{minipage}[c]{0.48\linewidth}
    \vspace{-20pt}
        \includegraphics[width=0.9\linewidth]{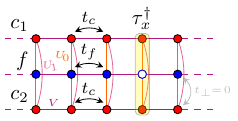}
    \end{minipage}
    \llap{
    \parbox{0pt}{\vspace{-92pt}\hspace{0pt}\footnotesize{(b)}}
    }
    \begin{minipage}[c]{0.48\linewidth}
        \includegraphics[width=1\linewidth]{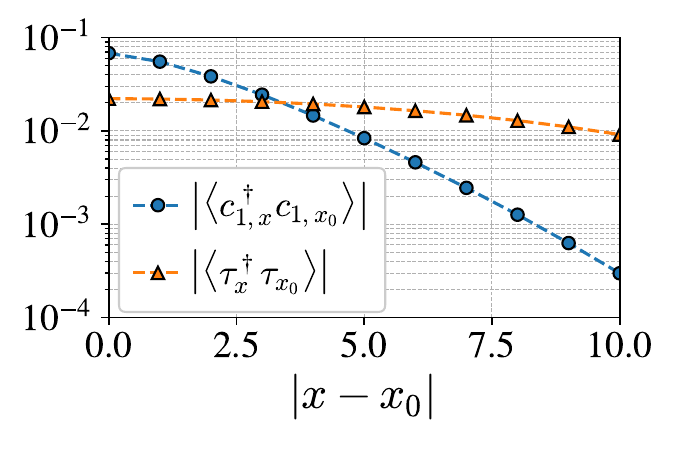}
    \end{minipage}

    \caption{(a) Schematic illustration of the model in Eq.~\eqref{eq:H2c1f}. The operator $\tau_x^\dagger=c^\dagger_{1,x} f_x c^\dagger_{2,x}$ is a trion creation operator at position $x$ along the chain.
    (b) Single-electron and trion static correlation functions in the trion liquid phase with parameters corresponding to the point marked by a star in Fig.~\ref{fig:PD} and $V/U_0=1/6$, in the presence of 5 trions in a system of $N=101$ sites ($\sim5\%$ electron doping). Correlations are plotted as a function of the distance from the center of the chain, $x-x_0$.} While single electrons are gapped, trions exhibit power-law correlations.
    \label{fig:model}
\end{figure}

As an explicit demonstration of the trion liquid phase, we study a minimal 1D microscopic model motivated by the crystal structure of \TNS. The model consists of three coupled chains, as depicted in Fig.~\ref{fig:model}(a). For simplicity, we neglect the spin degree of freedom and describe the system by the following spinless-fermions Hamiltonian
\begin{align}\label{eq:H2c1f}
    H &= H_0+H_{\rm int} \nonumber \\
    H_0 &= -t_c \sum_{\alpha,i} c^\dagger_{\alpha,i} c_{\alpha,i+1}
    +t_f \sum_{i}f^\dagger_{i} f_{i+1} \nonumber \\
    & +\frac{D}{2} \sum_{i} \Big( n^c_{i} - n^f_{i}\Big)
    -\mu \sum_{i} \Big( n^c_{i} + n^f_{i}\Big) \nonumber
    \\
    H_\text{int} &=  
    U_0 \sum_{i} n^c_{i} n^f_{i}
    + U_1 \sum_{i} n^c_{1,i} n^c_{2,i} \nonumber \\
    & + V \sum_{i} \Big( \sum_{\alpha} n^c_{\alpha,i} n^c_{\alpha,i+1} +n^f_{i} n^f_{i+1} \Big).
\end{align}
Here $c_{\alpha,i}$ is the annihilation operator of a conduction electron on site $i$ on chain $\alpha=1,2$, $f_{i}$ is the annihilation operator of a valence electron at site $i$. The respective $c_\alpha$ ($f$)-electron number operator is given by $n^c_{\alpha,i}$ ($n^f_i$), and $n^c_i=n^c_{1,i}+n^c_{2,i}$.
The hopping amplitudes $t_c$ and $t_f$, of the $c_\alpha$ and $f$ fermions, are taken to be positive, resulting in electron-like and hole-like bands, respectively. 
The size of the single particle gap is tuned by $D$
and $\mu$ is the chemical potential. The strength of the interaction between the nearest-neighbor conduction and valence electrons, $U_0$, determines the exciton binding energy, while the interaction between the conduction electrons on the two chains within the same unit cell, $U_1$, controls the trion binding. Finally, we take the nearest-neighbor intra-chain interaction, $V$, to be equal on all three chains. Its precise value does not affect our main findings; a small nonzero $V$ stabilizes the dilute trion liquid by inducing repulsion between trions and preventing larger bound complexes.

Note that in the model considered there is no hybridization between the $f$ and $c$ chains ($t_\perp=0$). As a result, the conduction and valence bands retain their orbital character, and the interactions take a purely density-density form in the band basis. 
The number of $c$-electrons and the number of $f$-electrons are then independently conserved. 
We also take the inter-chain $c$ hopping to be zero. While non-zero hopping would reduce $\Eexe$, our results remain qualitatively unchanged as long as the hopping is small compared to the binding energy.

We study the model numerically, using density matrix renormalization group (DMRG)~\cite{White1992,Schollwock2011} to obtain the phase diagram, and the time-dependent variational principle (TDVP)~\cite{HaegemanVerstraete2011,HaegemanVerstraete2016} to probe the dynamical correlations.
Calculations were performed using the ITensor library~\cite{itensor,itensor-r0.3}.
Results presented were obtained on a system with open boundary conditions of length $N=101$ unless stated otherwise.
 
\section{Phase Diagram}
We start by analyzing the model at charge neutrality, namely, for a fully occupied valence band and an empty conduction band. A similar analysis was employed in the past studying a closely related two-band model~\cite{Ejima2014}.
To obtain the phase boundary between the EI and BI phases, we calculate the exciton binding energy and the single-particle gap. Denoting by $E(N_c,N_f)$ the lowest energy state of the system with $N_c$ conduction electrons (in both c-chains), and $N_f$ valence electrons, the binding energy (defined to be positive) is given by 
\begin{equation}
    \Eex = -\left[ E(1,N\!-\!1)\! +\! E(0,N)\! -\! E(1,N)\! -\! E(0,N\!-\!1)\right],
\end{equation}
where $N$ is the number of unit cells in the system (at charge neutrality $N_c=0$, $N_f=N$).
The single-particle gap can be calculated as
\begin{equation}
    \Eg = E(1,N)\!+\!E(0,N\!-\!1)\!-\!2E(0,N),
\end{equation}
and is given explicitly by $\Eg=D - 2(t_c+t_f) + U_0 - 2V$ for the model considered.
A transition into the EI phase occurs when $\Eex>\Eg$. The phase diagram as a function of $D$ and $U_0$ obtained using this criterion is shown in Fig.~\ref{fig:PD}.

We note that in a 1D system, true long-range excitonic order is not possible due to phase fluctuations of the excitonic order parameter. However, the EI phase can still be distinguished by pseudo-long-range order (a power-law decay of the excitonic correlations) alongside a single particle gap.
Furthermore, upon a transition into the EI phase, the occupation of the $c$-electrons at charge neutrality becomes non-zero, $\expval{n_c}>0$.

We next analyze the model upon electron doping. To probe the regime in which trion formation is possible we calculate the trion binding energy
\begin{align}
    \Etr 
    &=
    \Eex+\Eexe
    \\\nonumber
    E^b_{\text{ex-e}}
    &=
    \!-\!\left[ E(2,N\!-\!1)\! +\!E(0,N)\! -\! E(1,N\!-\!1)\! -\! E(1,N)\right].
\end{align}

As discussed above, the trion liquid can form when $\Etr > \Eg$. Specifically, we consider the regime in which the trions form in the absence of an excitonic condensate $\Eex<\Eg$. This region, obtained for a fixed ratio of $U_1/U_0=1/3$, is indicated on the phase diagram in Fig.~\ref{fig:PD} with dashed lines.

\section{Photoemission spectrum in the trion liquid phase}
While the equilibrium properties of the negatively- and positively-charged trion liquids are similar, the photoemission spectrum in the two cases differs significantly.
Our focus here is on the distinct spectral signatures of negatively-charged trions, and we comment on the expected behavior for positively-charged trions in the discussion.

We calculate the photoemission single-particle spectral function, $A^-(k,\omega)$, defined as the imaginary part of the lesser Green’s function, considering the contributions from the $c$ and $f$ bands separately.
This quantity is experimentally accessible via, e.g., an ARPES measurement~\cite{Damascelli2004}.
We note that when the conduction and valence bands have distinct orbital characters, a band-resolved photoemission spectrum can be experimentally accessed by exploiting the orbital selectivity provided by photon polarization.

At zero temperature, 
\begin{align}\label{eq:Akw}
A^-_d(k,\omega) & =
\sum_m \left|\langle m| d_k |0\rangle\right|^2 \delta\left(\omega - (E_m-E_0)\right) \nonumber \\
& =\int dt e^{-i\omega t}\sum_j e^{ikj} \langle 0| d_j^\dagger (t) d_0(0) |0 \rangle.
\end{align}
Here $d$ stands for either the fermionic annihilation operator $c_\alpha$ or $f$, and the states  $\left|0\right\rangle$ and $\left|m\right\rangle$ denote the ground state and the $m$th excited state, respectively.
The full photoemission spectrum is given by $A^-(k,\omega) = A^-_c(k,\omega)+A^-_f(k,\omega)$,
where the contribution to the $c$-channel spectrum comes from both chains, $A^-_c=A^-_{c_1}+A^-_{c_2}$. (In the numerical results presented below, we calculate $A^-_{c_1}$, and assume $A^-_{c_2}=A^-_{c_1}$ due to symmetry.)

As mentioned above, to calculate the time-dependent correlations, we employ TDVP.
We use the 2-site TDVP algorithm, with a time step of $dt=0.3$, complemented with global subspace expansion~\cite{YangWhite2020} of 2nd order at each time step. Specifically, we obtain the quenched state $d_0\ket{0}$ (here the site index $j=0$ corresponds to a site in the middle of the chain) and evolve it in time.
The time evolution is carried out up to times $T=75$ and $T=25$ (in units of $t_c^{-1}$) for the $c$- and $f$-channel, respectively, making sure the portion of the light-cone that reaches the end of the chain is negligible. (Longer evolution times in the $c$-channel are possible due to the large effective mass of the trions, which dominate the $c$-channel response.) A maximal bond dimension of $200$ is used, resulting in truncation errors of order $10^{-5}$ at late times. To enhance the frequency resolution and avoid ringing effects due to the finite integration time, after the spatial Fourier transform, we perform linear prediction
followed by Gaussian windowing~\cite{White2008}.

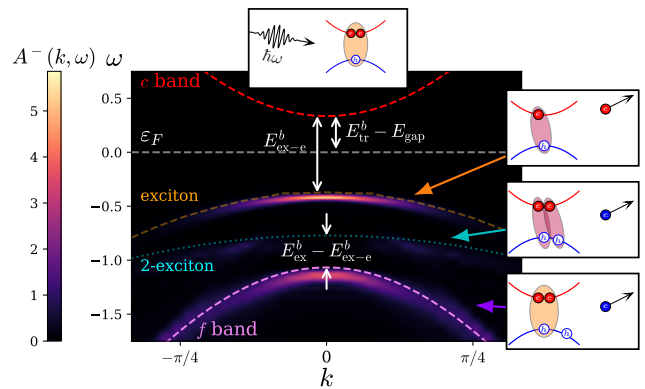
\begin{figure}[t]
    \centering
    \resizebox{\linewidth}{!}{
        \begin{tikzpicture}
        \definecolor{mpl-orange}{rgb}{1.0, 0.498, 0.055}
        \definecolor{mpl-cyan}{rgb}{0.0, 0.75, 0.75}
        \definecolor{mpl-violet}{rgb}{0.56, 0.0, 1.0}
        \tikzset{
            compactfig/.style={
                draw=black,
                draw opacity=1,
                fill=white,
                fill opacity=1,
                text opacity=1,
                inner sep=0pt,
                anchor=center,
                thick 
            }
        }
            \node at (0,0) {\includegraphics[width=1.0\linewidth]{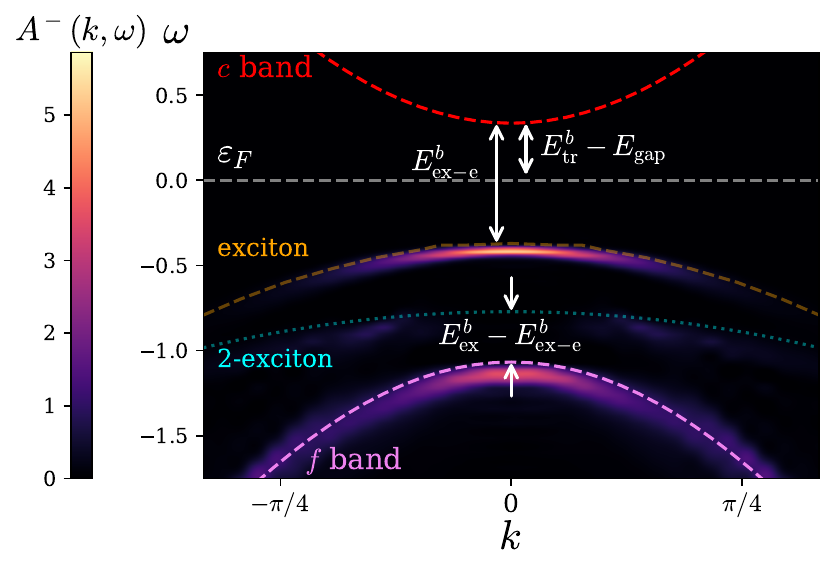}};
            \node[compactfig,align=center] at (1.0,2.8) {\includegraphics[width=0.3\linewidth]{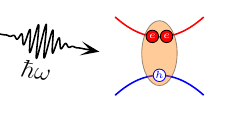}};

            \node[compactfig] (c_channel) at (5.0,1.5) {\includegraphics[trim={3 0 0 0}, clip,width=0.25\linewidth]{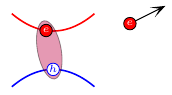}};
            
            \node[compactfig] (f1channel) at (5.0,0.0) {\includegraphics[trim={3 0 0 0}, clip,width=0.25\linewidth]{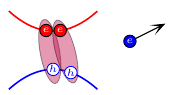}};
            
            \node[compactfig](f2channel) at (5.0,-1.5) {\includegraphics[trim={3 0 0 0}, clip,width=0.25\linewidth]{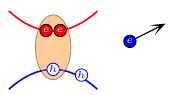}};
            
            \draw[mpl-orange,-{Latex[length=3mm]},line width=1] (c_channel) -- ++ (-2.6,-1.1);
            \draw[mpl-cyan,-{Latex[length=3mm]},line width=1] (f1channel) -- ++ (-2.0,-0.3);
            \draw[mpl-violet,-{Latex[length=3mm]},line width=1] (f2channel) -- ++ (-1.6, 0.1);
        \end{tikzpicture}}

    \caption{Photoemission spectrum, $A^-(k,\omega)=A^-_c(k,\omega)+A^-_f(k,\omega)$, obtained numerically 
    using model parameters corresponding to the point marked by a star in Fig.~\ref{fig:PD} and $V/U_0=1/6$, in the presence of 3 trions in a system of size $N=101$ sites ($\sim3\%$ electron doping). 
    Insets schematically depict the processes contributing to the spectral weight, with the emitted electron shown in red (blue) for processes in the $c$ ($f$) channel.
    The red (purple) dashed line corresponds to the non-interacting $c$ ($f$) band. Exciton dispersion as well as the boundary of the 2-exciton continuum obtained using ED are plotted as orange dashed and cyan dotted lines, respectively.
    } 
    \label{fig:ARPES}
\end{figure}

The full photoemission spectrum $A^-(k,\omega)$, obtained in the low trion density regime ($\sim3\%$ electron doping) for model parameters corresponding to the point marked by a star in Fig.~\ref{fig:PD}, is shown in Fig.~\ref{fig:ARPES}. 
Band-resolved photoemission spectrum, as well as the evolution of the spectrum with increasing trion density, are presented in \ref{app:SpectralFunction}.

Most notably, we observe a gapped high-intensity feature, originating in the $c$-channel.
We attribute this feature to the breakup of an equilibrium trion, leaving behind an exciton. 
To understand the expected spectral signature of such a process in the regime of low trion density, consider the case of a single trion. 
Upon photoemission of a $c$-electron, the system is left in an excited state containing an exciton, with energy $\Delta E=\Eg-\Eex$. Thus, finite spectral weight corresponding to this process is expected at energy $\Delta E$ below the Fermi energy. Using Eq.~\eqref{eq:mu_c} for the Fermi energy at which the first trion is formed, this corresponds to an energy $\Eexe$ below the bottom of the conduction band, consistent with the additional energy cost of breaking the trion compared with free-electron emission.

In the model considered here, the lowest-energy trions have zero center-of-mass momentum (see EM) and are composed predominantly of zero-momentum electrons and excitons. Thus, photoemission of an electron with momentum $k$ close to zero leaves behind an exciton with momentum $-k$, resulting in a feature tracing the exciton dispersion.
More generally, for a trion with nonzero center-of-mass momentum $k_{\rm tr}$, as may occur for example in two-dimensional TMDs, photoemission of an electron with momentum $k$ leaves a residual exciton with momentum $k_{\rm tr}-k$, and the largest spectral weight occurs at electron momenta that contribute most strongly to the trion wavefunction~\cite{Meneghini2025,wu2025signature}.

To confirm this interpretation, we calculate the spectrum within the 1-electron-1-hole subspace using exact diagonalization (ED) as detailed in Appendix ~\ref{app:ED}. The dispersion of the exciton obtained in ED is plotted at energy $\Eg-\Eex$ below the Fermi energy in Fig.~\ref{fig:ARPES} (see orange dashed line). We find an excellent agreement between the dispersion of the feature observed in the spectral function and the exciton dispersion.

Upon increasing electron doping (see Appendix ~\ref{app:SpectralFunction}), the intensity of the feature rises due to the increase in trion density. Furthermore, at higher trion density, trions start occupying higher-momentum states. Removing an electron from a trion with momentum $k_\tau$ produces peak photoemission intensity when the exciton left behind has zero COM momentum. This corresponds to a photoemission of an electron with momentum $k=k_\tau$. Consequently, the dispersion of this feature starts to resemble the trion band dispersion.  

In the $f$-channel, the dominant weight 
traces the $f$-band dispersion. In addition, we observe a weaker signal above the top of the valence band. 
We attribute this weight to a process in which 
the trion present in the system and the hole left behind rearrange to form a pair of excitons, gaining energy equal to $2\Eex-\Etr=\Eex-\Eexe$.
Calculating the 2-electrons-2-holes spectrum using ED, we indeed observe a 2-exciton continuum at energy $\Eex-\Eexe$ below the 1-trion-1-hole continuum (see Fig.~\ref{fig:ed} in the EM). The boundary of the 2-exciton continuum is plotted as a cyan dotted line in Fig.~\ref{fig:ARPES}.

\section{Discussion}
We comment on the spectral signatures expected in the photoemission spectrum for a positively charged trion liquid. 
Since the trion in this case is built from an electron and two holes, photoemission of a (conduction) electron will leave behind two holes. This will result in a broad two-hole continuum at energy $\Etr$ below the bottom of the conduction band. Since in the trion liquid regime $\Etr>\Eg$, this two-hole continuum will overlap with the valence band. However, its orbital character will be that of the conduction band, potentially allowing it to be distinguished experimentally via dependence on light polarization.

Although the microscopic model we considered in this work is a 1D model, we anticipate our main findings to extend to the 2D case. 
In particular, we anticipate the emergence of an in-gap state in photoemission for negatively charged trions, with vanishing spectral weight at the Fermi level, although the system hosts mobile charge carriers. Furthermore, we expect probes of the density of states to reveal finite spectral weight within the single-particle gap, for both negatively and positively charged trions. A detailed analysis of these processes is left for future work.

\begin{acknowledgments}
We thank Amit Kanigel, Yuval Nitzav,  Hadas Soifer, and Ittai Sidilkover for fruitful discussions.
A.K.\ acknowledges funding by the Israel Science Foundation (Grant No.\ 2443/22).
\end{acknowledgments}

\appendix
\section{Spectral function upon increasing trion density}\label{app:SpectralFunction}

\begin{figure}[b]
    \centering
    \includegraphics[width=\linewidth]{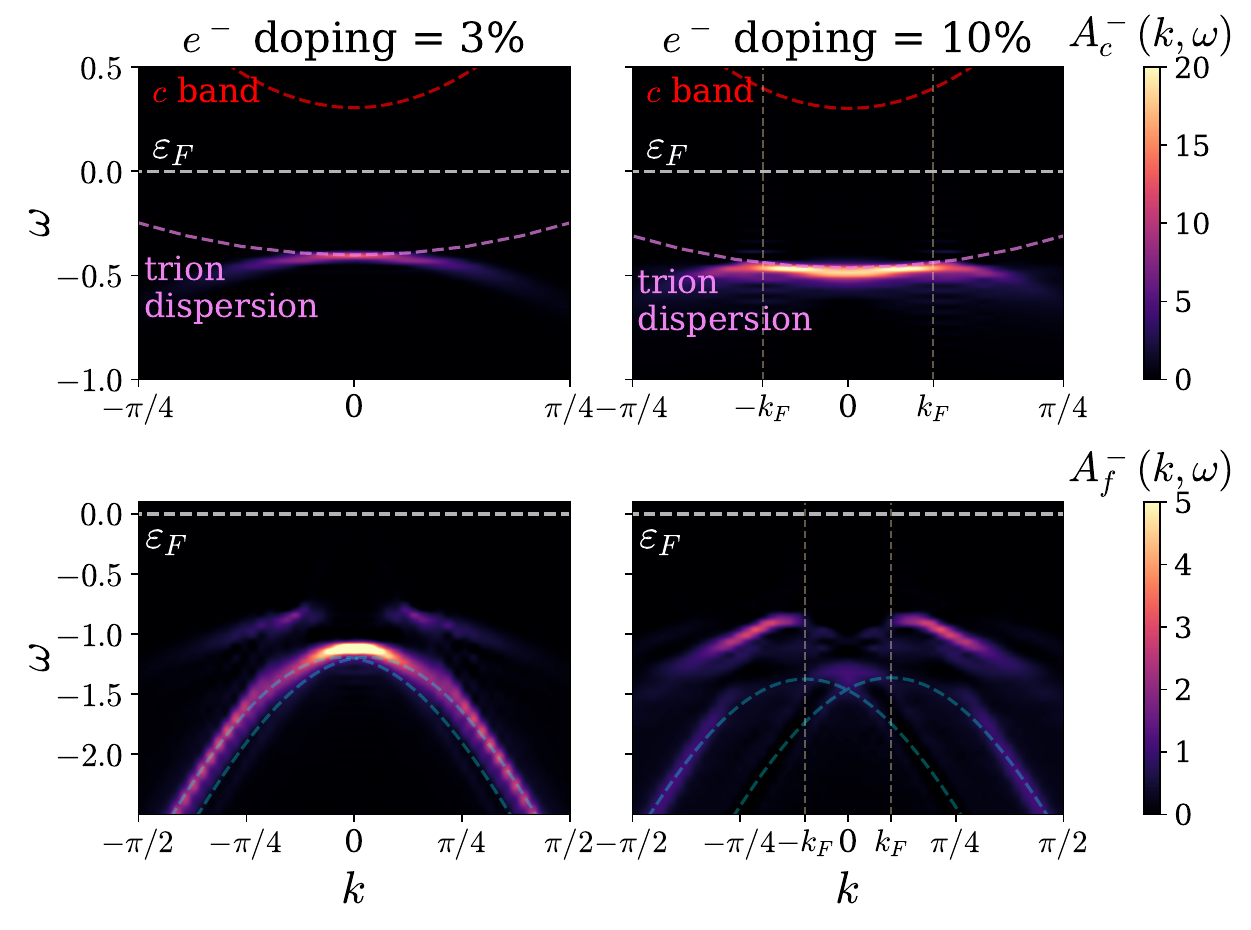}
    \caption{Band-resolved photoemission spectrum at $3\%$ and $10\%$ electron doping. With increasing trion density the in-gap feature in the $c$-channel evolves to have the trion band dispersion (vertical grey dashed lines indicate the Fermi momenta of the trions). In the $f$-channel a $\pm \kF$ splitting of the valence band is observed, originating from the fermionic commutation relations between the trions and the $f$-electrons. The intensity of the in-gap two-exciton feature also increases.}
    \label{fig:high_dopping_spectral_fun}
\end{figure}

In Fig.~\ref{fig:high_dopping_spectral_fun} we show band-resolved photoemission spectra, $A^-_{c,f}(k,\omega)$, both in the low trion density regime ($\sim 3\%$ electron doping as in Fig.~\ref{fig:ARPES} in the main text), and for a higher trion density ($\sim10\%$ electron doping). As can be seen, when trion density is increased, the intensity of the feature in the $c$-channel increases, and its dispersion evolves from a hole-like dispersion with spectral weight centered around $k=0$, to an electron-like one, with a large effective mass. Obtaining the dispersion of the trion band using ED (see below) we find a very good agreement between the trion dispersion and that of the in-gap feature. Furthermore, the momentum-space region in which the feature exhibits finite spectral weight matches well the expected occupation of the trion band (namely, corresponding to momentum states up to $\kF$ as determined by the trion density).

In the $f$-channel, we observe a splitting of the non-interacting $f$-band, proportional to the density of trions. This is due to the 1D nature of the problem, and the fermionic commutation relations between the $f$-electrons and the trions~\cite{Feiguin2009}.
In addition, we observe an increase in the intensity of the feature appearing above the top of the valence band. This is again consistent with the expectation that for a larger trion density the amplitude for a process in which a hole and a trion recombine to form a pair of excitons is increased.

\section{Exact diagonalization spectra}\label{app:ED}

\begin{figure}
    \centering
    \begin{minipage}[c]{1\linewidth}
        \includegraphics[width=\linewidth]{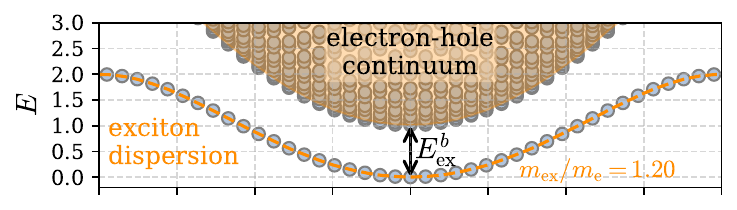}
    \end{minipage}
    \llap{
    \parbox{0pt}{\vspace{-135pt}\hspace{-220pt}\footnotesize{(a)}}
    }
    \begin{minipage}[c]{1\linewidth}
        \includegraphics[width=\linewidth]{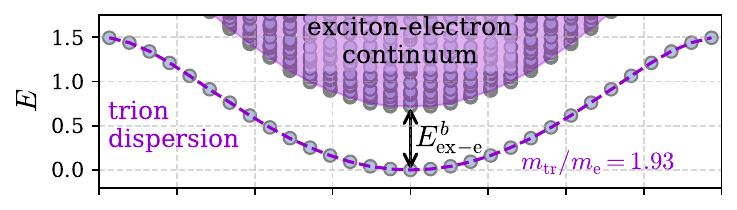}
    \end{minipage}
    \llap{
    \parbox{0pt}{\vspace{-140pt}\hspace{-220pt}\footnotesize{(b)}}
    }
    \begin{minipage}[c]{1\linewidth}
        \includegraphics[width=\linewidth]{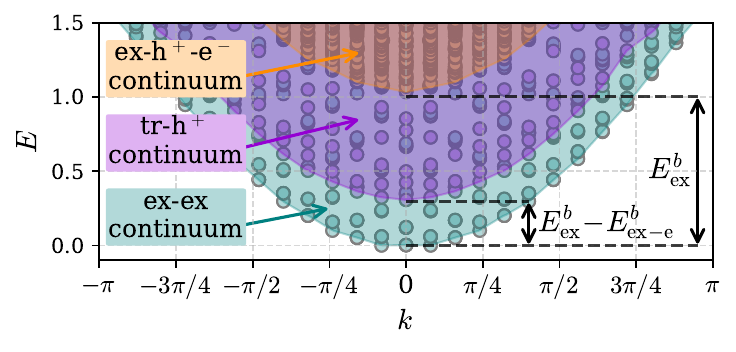}
    \end{minipage}
    \llap{
    \parbox{0pt}{\vspace{-230pt}\hspace{-220pt}\footnotesize{(c)}}
    }
    
    \caption{The low-energy spectrum obtained using ED on a system with periodic boundary conditions and model parameters corresponding to the point marked by a star in Fig.~\ref{fig:PD} ($U_0=3.0,\;U_1=1.0,\;V=0.5$). 
    (a) The 1-electron-1-hole spectrum for an $N=41$-long chain. The low energy band separated from the continuum gives the exciton dispersion.
    (b) The 2-electron-1-hole spectrum for an $N=31$-long chain. The low energy band separated from the continuum gives the trion dispersion.
    The binding energies in (a) and (b) are in excellent agreement with the values obtained in DMRG.
    (c) The 2-electron-2-hole spectrum for an $N=25$-long chain. The 2-exciton continuum (shown in green) overlaps with both the 1-trion-1-hole continuum (shown in violet) and the 1-exciton-1-hole-1-electron continuum (shown in orange). The bottom of the two continua are at $\Eex-\Eexe$, and $\Eex$, above the bottom of the 2-exciton continuum, respectively.
    }
    \label{fig:ed}
\end{figure}

Performing a particle-hole transformation on the $f$-electron operators in the Hamiltonian in Eq. \eqref{eq:H2c1f} allows us to obtain the low-energy spectrum in the vicinity of charge neutrality efficiently using exact diagonalization (ED).
Specifically, using periodic boundary conditions, we calculate the spectra within the sectors with 1-electron-1-hole, 2-electrons-1-hole, and 2-electrons-2-holes on systems of size $N=41$, $N=31$, and $N=25$ respectively. These are presented in Fig.~\ref{fig:ed} for model parameters corresponding to the point marked by a star in Fig.~\ref{fig:PD} ($t_c=t_f=1$, $U_0=3.0$, $U_1=1.0$) and $V=0.5$.

In the 1-electron-1-hole spectrum, the appearance of the electron-hole (exciton) bound state is clearly visible at energy $\Eex$ below the electron-hole continuum. 
Similarly, in the 2-electron-1-hole spectrum, the exciton-electron (trion) bound state appears at energy $\Eexe$ below the continuum. The magnitude of the binding energies is in excellent agreement with the DMRG calculation.

In the 2-electron-2-hole continuum, we identify the 2-exciton and the 1-trion-1-hole continua from the spatial structure of the eigenstates. As expected, the bottom of the 2-exciton continuum is at $2\Eex-\Etr=\Eex-\Eexe$ below the bottom of the 1-trion-1-hole continuum.

\bibliographystyle{unsrt}
\bibliography{refs}

\end{document}